\documentclass{article}
\usepackage{graphicx}
\usepackage[english]{babel}
\usepackage[margin=1in]{geometry}
\usepackage{multicol}
\usepackage{subcaption}
\author{S. Chatterjee \footnote{Indian Insitute of Information Technology, Allahabad \texttt{$\lbrace$rs171@iiita.ac.in$\rbrace$}} \and B.S. Sanjeev \footnote{Indian Insitute of Information Technology, Allahabad \texttt{$\lbrace$sanjeev@iiita.ac.in$\rbrace$}} }
\title{
{\bfseries\Large Role of Toll-Like Receptors in the interplay between pathogen and damage associated molecular patterns\bigskip}
}

\date{July 1, 2019}

\begin{document}
\maketitle

\begin{center}
Department of Applied Sciences\\ Indian Institute of Information Technology\\ Allahabad 211012, India \\
\end{center}

%S. Chatterjee\textsuperscript{1} and B.S. Sanjeev\textsuperscript{1} \\

%
%\normalfont (Dated: July 1, 2016)

\tableofcontents

%\newpage

\begin{abstract}
%Insert your abstract here. Include keywords, PACS and mathematical subject classification numbers as needed.

Various theoretical studies have been carried out to infer relevant protein-protein interactions among pathogens and their hosts. Such studies are generally based on preferential attachment of bacteria / virus to their human receptor homologs. We have analyzed 17 pathogenic species mainly belonging to bacterial and viral pathogenic classes, with the aim to identify the interacting human proteins which are targeted by both bacteria and virus specifically. Our study reveals that the TLRs play a crucial role between the pathogen-associated molecular patterns (PAMPs) and the damage associated molecular patterns (DAMPS). PAMPs include bacterial lipopolysaccharides (LPs), endotoxins, bacterial flagellin, lipoteichoic acid, peptidoglycan in bacterial organisms and nucleic acid variants associated with viral organisms, whereas DAMPs are associated with host biomolecules that perpetuate non-infectious inflammatory responses. We found out the presence of SOD1 and SOD2 (superoxide dismutase) is crucial, as it acts as an anti-oxidant and helps in eliminating oxidative stress by preventing damage from reactive oxygen species. Hence, such strategies can be used as new therapies for anti-inflammatory diseases with significant clinical outcomes.

%\keywords{Molecular Networks \and Pathogen-Host Interactions \and Important Interacting Proteins \and Centrality Indices}

% \PACS{PACS code1 \and PACS code2 \and more}
% \subclass{MSC code1 \and MSC code2 \and more}
\end{abstract}

\section{Introduction}
\label{intro}
%Your text comes here. Separate text sections with

The interactions between pathogens (e.g. virus, bacteria, etc.) and their host (e.g. humans, plants) can be illustrated on a single-cell level (individual encounters of pathogen and host), on a molecular level (e.g. pathogenic protein binds to receptor on human cell), at the level of an organism (e.g. virus infects host), or on the population level (pathogen infections affecting a human population). Investigating the host and the pathogen protein networks in the host-pathogen interactome may allow us to better understand the crucial intermediaries in action which aid in the functioning of the host immune system. The important interacting proteins present in the host cell could be utilized as potential drug targets~\cite{1}. A system of interacting elements can be examined using graph theoretical techniques.

\section{Materials and Methods}
\label{sec:1}

We make use of centrality indices to identify the most important vertices within a graph~\cite{3}$^,$~\cite{4}. Herein, for our undirected graph $\Gamma$, such that $\Gamma_p$ = ($V_p$, $E_p$), the set of all nodes (protein coding genes) is denoted by ($V_p$). $E_p$ is the set of corresponding edges (interactions) between human proteins ($H_p$) and pathogen organisms ($P_p$). The important interacting proteins in the constructed human host-pathogen bipartite graph were listed as hubs and subsequently studied the extent of their vulnerability to the pathogenic proteins. We have used the DisGeNET database (v 4.0)~\cite{2} and cross-referenced the pathogens that are specifically linked with diseases associated with infectious pathogens. The linked pathogens are either directly responsible to cause the infectious disease or are known to be found frequently in patients suffering from the disease. The entries in the dataset are manually curated and also supported by strong experimental evidence.

Our curated dataset linked with 17 pathogenic organisms comprises of 7,423 number of associations (edges) that interact with 2,648 unique human protein coding genes (nodes).

\subsection{Network construction from Data}

\begin{table}
\centering
\caption{{Number of Pathogens and their Class (Bacteria/ Virus).}\label{table:Host_Pathogen_Interactome_Data}}
{\begin{tabular}{llll}
\hline
Sl.	& Pathogens	& Class	& Degree($\Gamma$) \\
\hline
1 & Helicobacter pylori			& Bacteria  & 1150 \\
2 & Chlamydia pneumoniae 		& Bacteria  & 745 \\
3 & Borrelia 					& Bacteria  & 660 \\ 
4 & Toxoplasma gondii 			& Bacteria 	& 378 \\
5 & Streptococcus 				& Bacteria  & 109 \\
6 & Mycobacterium tuberculosis 	& Bacteria	& 72 \\
7 & Bartonella					& Bacteria 	& 56 \\
\hline
8 & Enteroviruses	 			& Virus 	& 1025 \\
9 & Cytomegalovirus	 			& Virus 	& 865 \\
10 & Epstein-Barr virus 		& Virus		& 477 \\
11 & Herpes simplex virus 		& Virus 	& 404 \\
12 & Parvovirus B19				& Virus 	& 390 \\
13 & Human herpesvirus 6 		& Virus 	& 386 \\
14 & Influenza A 				& Virus 	& 224 \\
15 & Hepatitis C virus	 		& Virus 	& 199 \\
16 & HIV 						& Virus 	& 180 \\
17 & Hepatitis B virus			& Virus 	& 103 \\
\hline
\end{tabular}}{}
\end{table}

\begin{figure}
\centering
\includegraphics[width=\textwidth]{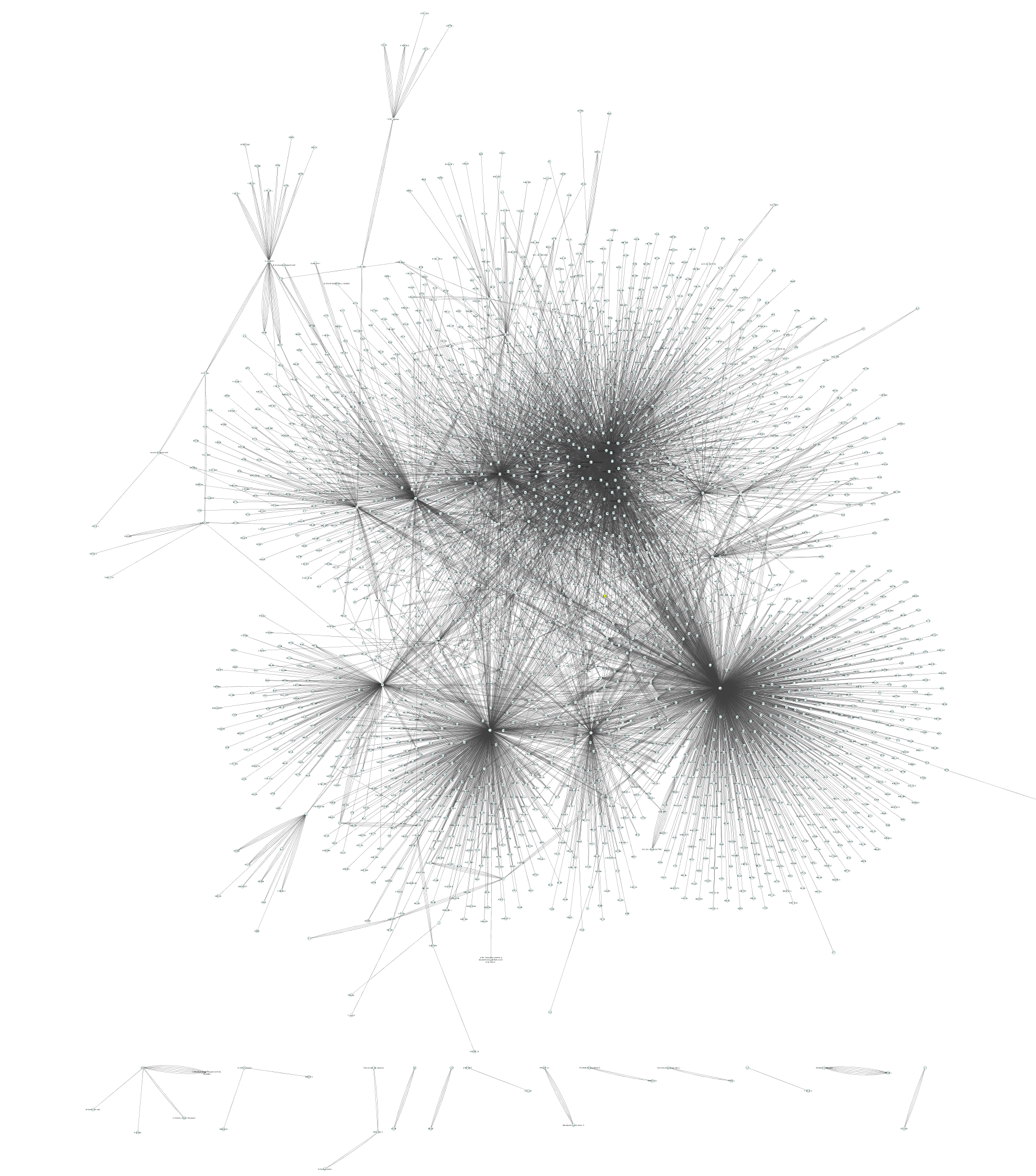}
\caption{The Pathogen-Host Network depicting the interactions (edges) between human and pathogenic proteins (nodes). Herein we can see the dense cluster of pathogenic genes around important interacting human proteins.}
\label{N_C_D}
\end{figure}

\begin{table}
\centering
\caption{Overview of PHI Networks}
\label{table:Analysis_of_PHI_Interactions}
\begin{tabular}{ll}
\hline
Parameters 												& Nos.(\#) \\
\hline
Total no. of pathogens									& 17 \\
No. of human proteins ($H_p$)							& 2,536	\\
Total no. of associations interacting with pathogens	& 7,019	\\
\hline
\end{tabular}
\end{table}

\subsubsection{Human-Pathogen Interactome Network}
We construct network based graph from the data that is experimentally validated based on physical interactions between the two distinct set of proteins. 

\subsection{Network Metrics}
Biological processes can be analyzed by using a wide scale of network based approaches in which the biological entities (e.g. genes, proteins or diseases) are represented in the form of nodes and edges which represent the type of interaction between such entities. Key nodes that are connected to multiple number of edges are known as hubs. A bipartite network is a graph in which edges connect a distinct set of source nodes to another set of target nodes (i.e. nodes that do not belong to the same set of nodes). Such approaches have been conceived theoretically and have been found useful in biomedical applications.

\begin{table}
\centering
\begin{tabular}{ll}
\hline
{\bf Parameter} 				& {\bf Statistics}	\\
\hline
Connected Components			& 13				\\
Network Diameter				& 9					\\
Network Centralization			& 0.264				\\
Shortest Paths					& 6,856,592 (97\%) 	\\ 
Characteristic Path Length		& 3.709				\\
Average number of Neighbours	& 3.681 			\\
Network Heterogeneity			& 6.519				\\
Multi-edge node pairs			& 1324				\\
\hline
\end{tabular}
\caption{Summary of Network Statistics of PHI Networks}
\label{table:Summary_of_PHI_Networks}
\end{table}

\section{Results and Discussion}

\subsection{Node Degree Distribution}
%{\bf Definition:}
The node degree distribution is an indicator of the number of nodes with a degree of {\it $k$}. The node degree of a particular node {\it $n$} is the number of interacting edges linked to {\it $n$}, in undirected networks.

\begin{figure}[!htbp]
\centering
\includegraphics[width=3.5in]{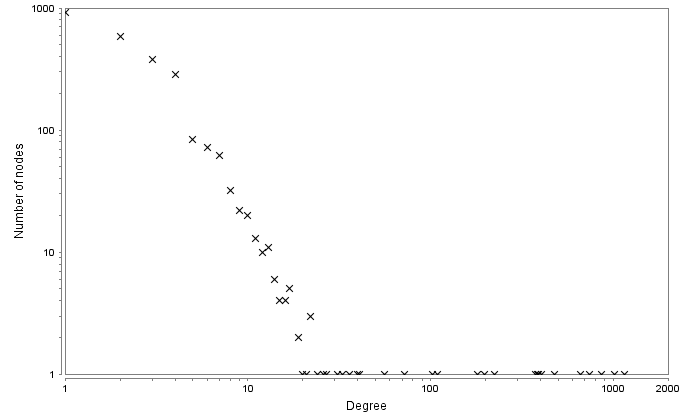}
\caption{Node Degree Distribution depicts the scale-free network topology of our pathogen-host network.}
\label{N_D_D}
\end{figure}

\subsection{Topological Coefficient}
%{\bf Definition:}
Topological coefficient {\it $T(n)$} of a node {\it $n$} with {\it $k_n$} neighbors is defined as the number of neighbors shared between a pair of nodes, {\it $k_n$} and {\it $k_m$}, divided by the number of neighbors of node {\it $k_n$}:

\begin{equation}
T(n)={\frac {avg(J(n,m))}{k_n}}
\end{equation}

where the value of {\it $J(n,m)$} is the number of neighbors shared between the nodes {\it $n$} and {\it $m$}, plus one if there is a direct link between {\it $n$} and {\it $m$}. It gives a relative measure of a node that shares it's neighbors with other nodes and is defined for all nodes {\it $m$} that share at least one neighbor with {\it $n$}.

\begin{figure}[!htbp]
\centering
\includegraphics[width=3.5in]{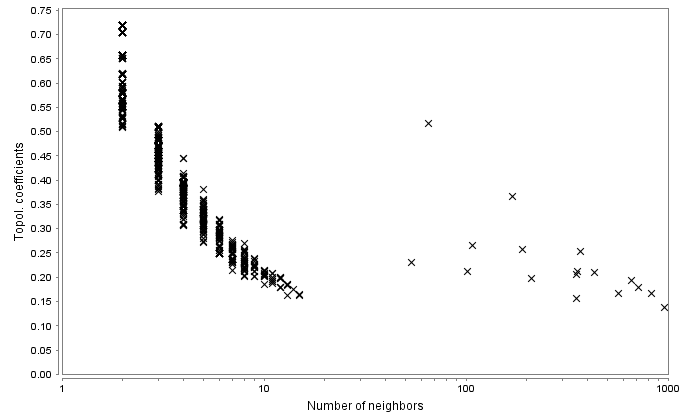}
\caption{Topological Coefficient is used to estimate the tendency of the nodes to have shared neighbors in the pathogen-host network.}
\label{T_C}
\end{figure}

\subsection{Betweenness Centrality}

Betweenness centrality of a node $v$ is the sum of the fraction of all-pairs shortest paths that pass through $v$.

\begin{equation}
C_B(v) =\sum_{s,t \in V} \frac{\sigma(s,t|v)}{\sigma(s,~t)}
\end{equation}

where $V$ is the set of nodes, $\sigma(s,t)$ is the number of shortest $(s,~t)$ paths, and $\sigma(s,t|v)$ is the number of those paths passing through some node $v$ other than $s,t$~\cite{5}. Hubs that tend to have high betwenness centrality are expected to lie in between many shortest paths and exhibit that even low-degree nodes with high betweeness may reveal a modular network structure~\cite{6}.

\begin{figure}[!htbp]
\centering
\includegraphics[width=3in]{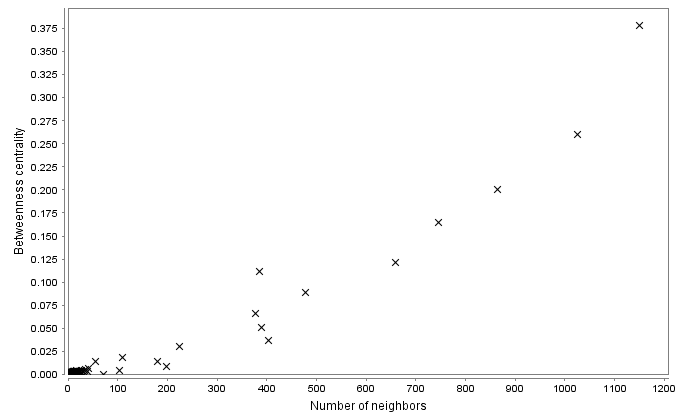}
\caption{Betweenness Centrality}
\label{B_C}
\end{figure}

\begin{figure*}
\centering
	\begin{subfigure}[b]{1.0in}
            \includegraphics[width=\textwidth]{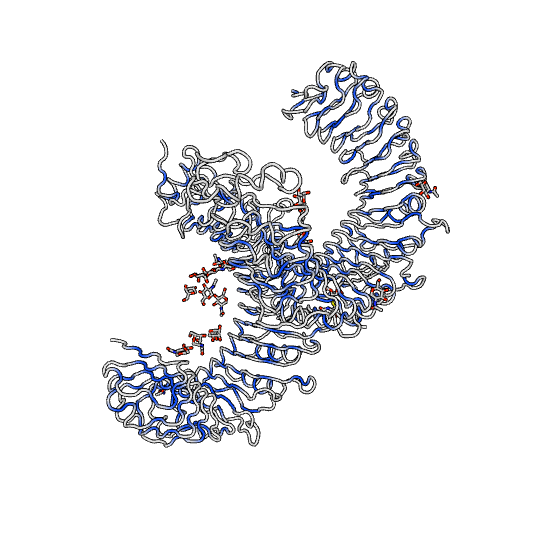}
            \caption{TLR1}
            \label{fig:TLR1}
    \end{subfigure}    
      %add desired spacing between images, e. g. ~, \quad, \qquad etc.
      %(or a blank line to force the subfigure onto a new line)
	\begin{subfigure}[b]{1.0in}
			\includegraphics[width=\textwidth]{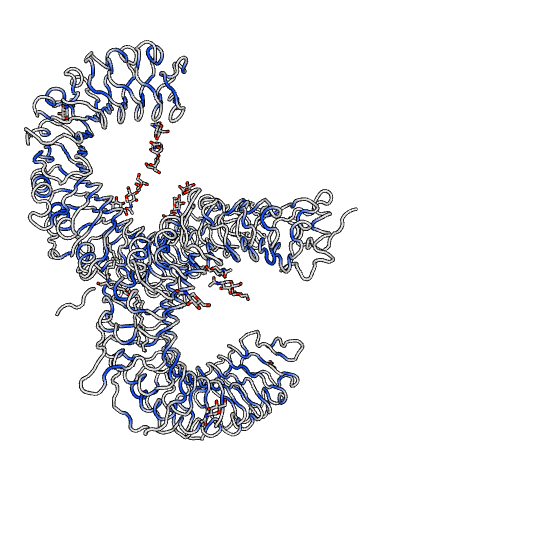}
            \caption{TLR2}
            \label{fig:TLR2}
    \end{subfigure}    
      %add desired spacing between images, e. g. ~, \quad, \qquad etc.
      %(or a blank line to force the subfigure onto a new line)    \begin{subfigure}[b]{1.2in} 
      \begin{subfigure}[b]{1.0in}           
            \includegraphics[width=\textwidth]{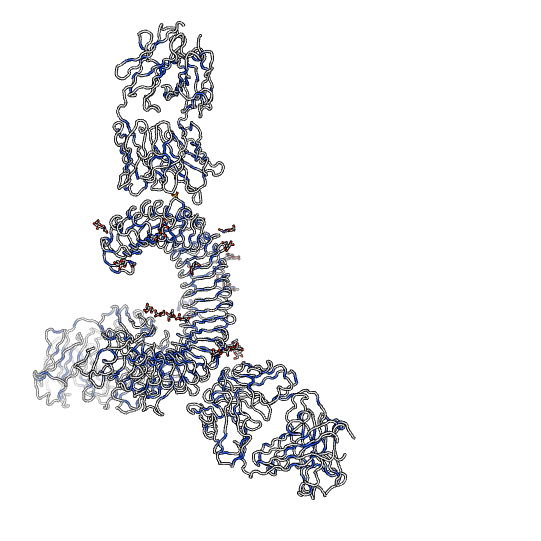}
            \caption{TLR3}
            \label{fig:TLR3}
    \end{subfigure}
      %add desired spacing between images, e. g. ~, \quad, \qquad etc.
      %(or a blank line to force the subfigure onto a new line)
    \begin{subfigure}[b]{1.0in}            
            \includegraphics[width=\textwidth]{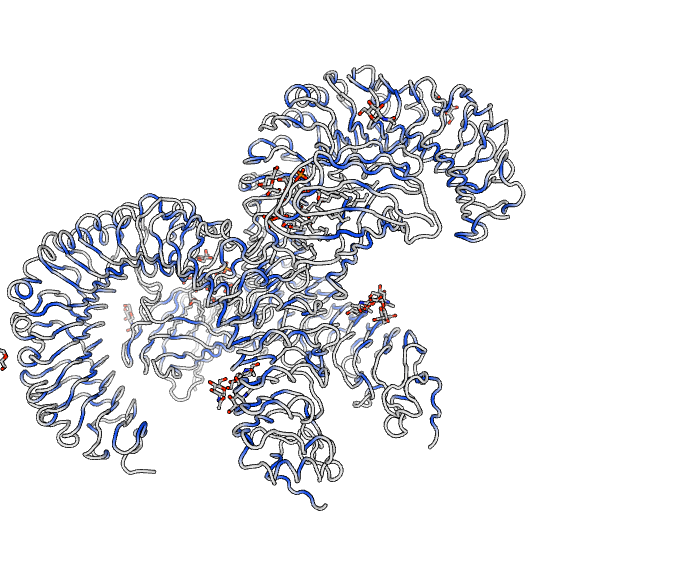}
            \caption{TLR4}
            \label{fig:TLR4}
    \end{subfigure}
      %add desired spacing between images, e. g. ~, \quad, \qquad etc.
      %(or a blank line to force the subfigure onto a new line)
    \begin{subfigure}[b]{1.0in}            
            \includegraphics[width=\textwidth]{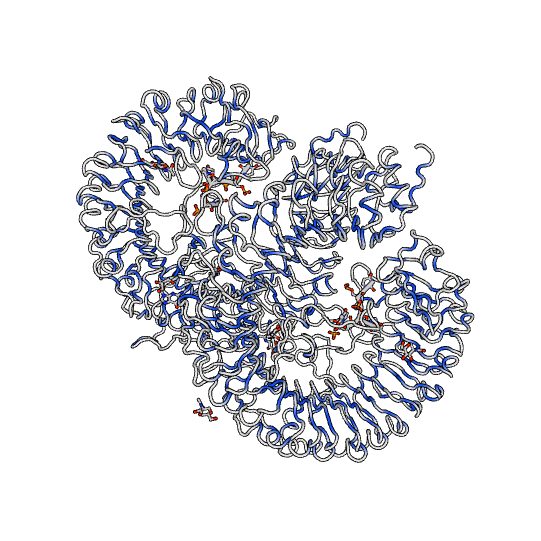}
            \caption{TLR7}
            \label{fig:TLR7}
    \end{subfigure}
      %add desired spacing between images, e. g. ~, \quad, \qquad etc.
      %(or a blank line to force the subfigure onto a new line)
    \begin{subfigure}[b]{1.0in}            
            \includegraphics[width=\textwidth]{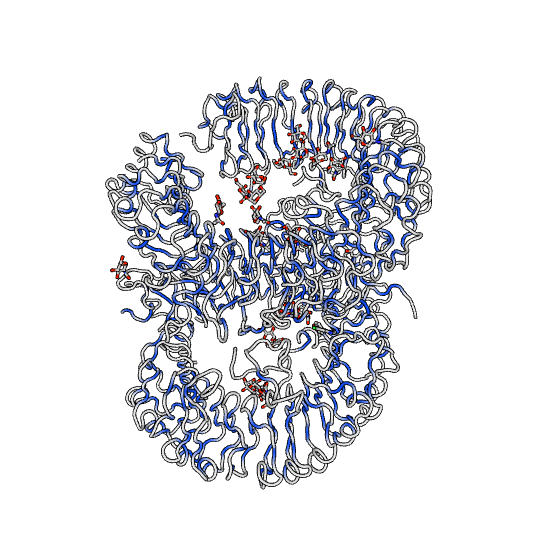}
            \caption{TLR8}
            \label{fig:TLR8}
    \end{subfigure}
    \caption{The blue-colored regions depict Leucine Rich Repeat regions (LRRs) of Toll-like receptors. Based on the number of interactions, Toll-like receptors \textit{viz.} TLR1, TLR2 and TLR4 were found to be more susceptible to bacterial proteins, whereas TLR3, TLR7 and TLR8 were more preferential to viral proteins. TLRs such as TLR5 and TLR6 were also found to be slightly more susceptible towards bacterial pathogenic proteins.}\label{fig:TOF}
\end{figure*}

\begin{table*}
\centering
\begin{tiny}
{\begin{tabular}{llllll}
\hline
Gene	& Degree	& Topol. Coeff. & Betw. Centrality. & Protein & Molecular Function\\
\hline
SOD2	& 41	& 0.16262985 & 0.00602848 & Superoxide dismutase [Mn], mitochondrial 			& Oxidoreductase 									\\
PTGS2	& 40	& 0.18507808 & 0.00330486 & Prostaglandin G/H synthase 2 						& Dioxygenase, Oxidoreductase, Peroxidase 			\\
TNF		& 36	& 0.16444268 & 0.00484062 & Tumor necrosis factor 								& Cytokine activity 								\\
TP53	& 33	& 0.1629871  & 0.00404872 & Cellular tumor antigen p53  						& Apoptosis, DNA-binding, Host-virus interaction  	\\
PLAU	& 31	& 0.19851805 & 0.0028075  & Urokinase-type plasminogen activator 				& Hydrolase, Protease, Serine protease 				\\
NOS2	& 27	& 0.1847816  & 0.00412495 & Nitric oxide synthase, inducible  					& Calmodulin-binding, Oxidoreductase 				\\
PLAT	& 26	& 0.2231064  & 0.00163375 & Tissue-type plasminogen activator  					& Hydrolase, Protease, Serine protease 				\\
IL6		& 24	& 0.19647329 & 0.00346007 & Interleukin-6  										& Cytokine, Growth factor 							\\
ACE		& 22	& 0.20865038 & 0.00227199 & Angiotensin-converting enzyme  						& Carboxypeptidase, Hydrolase, Protease 			\\
IL1B	& 22	& 0.19851805 & 0.0028075  & Interleukin-1 beta 									& Cytokine, Pyrogen, Inflammatory response 			\\
IFNG	& 22	& 0.19776016 & 0.00359721 & Interferon gamma 									& Antiviral defense, Growth regulation 				\\
SOD1	& 21	& 0.1748717  & 0.0034557  & Superoxide dismutase [Cu-Zn], soluble 				& Antioxidant, Oxidoreductase 						\\
STAT3	& 20	& 0.19647329 & 0.00346007 & Signal transducer and activator of transcription 3	& Activator, DNA-binding, Host-virus interaction 	\\
MTHFR	& 19	& 0.19866054 & 0.00260592 & Methylenetetrahydrofolate reductase  				& Allosteric enzyme, Oxidoreductase  				\\
CCL2	& 19	& 0.18357301 & 0.003561   & C-C motif chemokine 2 								& Cytokine , Chemotaxis, Inflammatory response 		\\
%16 & VEGFA  & 13	& 0.23189591 & 0.00229555 \\
\hline
\end{tabular}}
\end{tiny}
\caption{{Common Sub-group of Human Proteins during the cross-talk of pathogen and damage associated molecular patterns.}
\label{table:Gene_Pathogen_Dataset}}
\end{table*}

\begin{figure*}
\centering
\includegraphics[width=\textwidth]{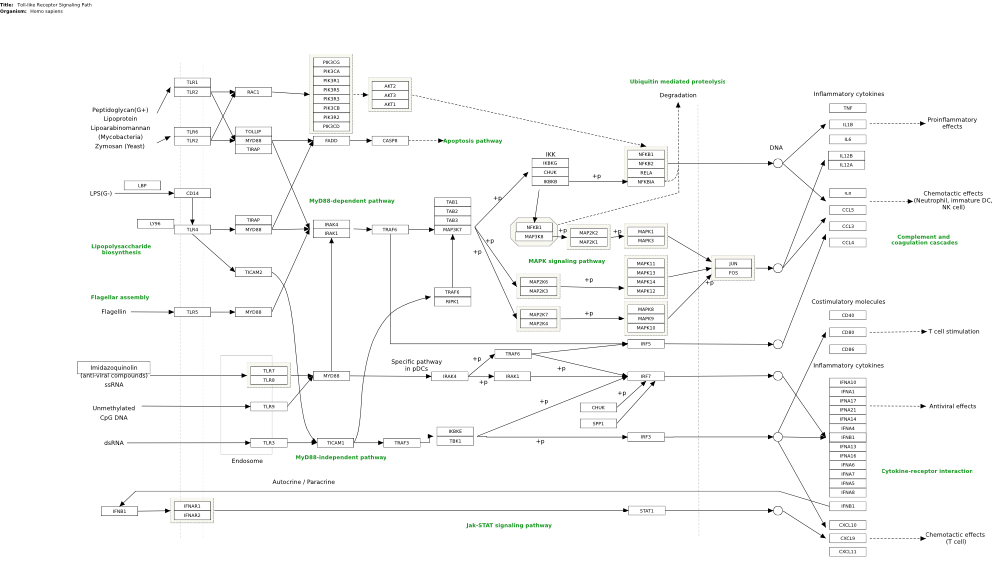}
\caption{Toll-like receptor (TLR) Signaling Pathway are membrane-bound host (human) proteins that respond to Gram-positive and Gram-negative bacteria. They induce pro-inflammatory cytokines and are responsible for producing innate immune responses.}
\label{S_P_L}
\end{figure*}

\subsection{Pathogen-associated molecular patterns}

The immune responses in response to the pathogen-associated molecular patterns (PAMPs) help protecting the host from infection. In our case, we found that the leucine rich-repeat (LRR) regions of Toll-like receptors initiate / perpetuate a cascade of mechanisms within the host, once they are able to detect the lipopolysaccharides(LPSs) and other  glycoconjugates found predominantly in bacteria and other nucleic acid derivatives in viruses. In our findings, we found that the TLRs frequently interact with the common sub-group of Human proteins that have higher values of Betweenness Centrality as depicted in Table~\ref{table:Gene_Pathogen_Dataset}. It has also been reported that the TLRs have a role in stimulating TNF and IL6 production~\cite{8}.

\subsection{Damage-associated molecular pattern}

Release of intra-cellular material leads to interfere with the membrane integrity and the ability to stimulate inflammatory response. DAMPs also have been found to play a crucial role to stimulate pattern-recognition receptors (PPRs) that comprise of Toll-likereceptors (TLRs) and leucine-rich repeat containing molecules. Thereafter, alarmins such as interleukins and interferons that are stored in cells and released upon cell lysis, are responsible for inflammatory responses.

\begin{table*}[!htbp]
\centering
\begin{tiny}
\begin{tabular}{llllll}
\hline
{\bf Gene} & {\bf Protein} & {\bf Uniprot ID} & \multicolumn{2}{c}{\bf No. of Pathogens} & \hspace{2.5cm} {\bf GO Molecular Function}\\
\cline{4-5} &&&{\bf Bacteria} & {\bf Virus} \\
\hline
TLR1	& Toll-like receptor 1	& Q15399	&	30	&	4	& TLR2 binding, Transmembrane signaling receptor activity\\
TLR2	& Toll-like receptor 2	& O60603	&	83	&	18	& Lipopolysaccharide receptor activity, Signaling PRR activity\\
TLR3	& Toll-like receptor 3	& O15455	&	8	&	29	& ds RNA binding, Transmembrane signaling receptor activity\\
TLR4	& Toll-like receptor 4	& O00206	&	63	&	22	& Lipopolysaccharide and Transmembrane signaling receptor activity\\
TLR5	& Toll-like receptor 5	& O60602	&	13	&	5	& Interleukin-1 receptor binding\\
TLR6	& Toll-like receptor 6	& Q9Y2C9	&	19	&	3	& Identical protein binding, lipopeptide binding, TLR2 binding\\
TLR7	& Toll-like receptor 7	& Q9NYK1	&	4	&	16	& ds/ss RNA binding, siRNA binding\\
TLR8	& Toll-like receptor 8	& Q9NR97	&   6	&	13	& ds/ss DNA binding, DNA binding, signaling receptor activity\\
TLR9	& Toll-like receptor 9	& Q9NR96	&	20	&	20	& interleukin-1 receptor binding, siRNA binding, \\
TLR10	& Toll-like receptor 10	& Q9BXR5	&	2	&	2	& Identical protein binding, Transmembrane signaling receptor activity\\
\hline
\end{tabular}
\end{tiny}
\caption{Overview of Toll-like receptors (TLRs) and their interactions with bacterial and viral proteins. The number depicted above signifies the number of interactions with bacterial/viral pathogenic proteins.}
\label{table:Analysis_of_IIPs}
\end{table*}

\section{Conclusions}

The predominant proteins/ protein coding genes which were found to indicate extensive cross-talk among themselves are described as follows:

The Superoxide dismutase [Mn], mitochondrial (SOD2) gene encodes an enzyme (Mn dependent) in humans which converts superoxide anion free radicals produced within cells to hydrogen peroxide and diatomic oxygen, thereby inhibiting cellular oxidative stress leading to apoptosis. As a result this gene eliminates the production of mitochondrial reactive oxygen species (ROS) and protects from programmed cell death. Interestingly, this gene has also been found to interact with p53 and NF$\kappa$B1 gene~\cite{9} which have earlier been found vulnerable to multiple pathogens.

The Prostaglandin-endoperoxide synthase 2 (PTGS2) gene encodes an enzyme which is responsible for the conversion of arachidonic acid to prostaglandin H2, that acts as a precursor of prostacyclin, and expressed in inflammation~\cite{10}. The PTGS2 gene is generally unexpressed under typical conditions but the levels are up-regulated during inflammation. The present line of treatment lies in selective inhibition of PTGS1 (COX-1) and PTGS2 (COX-2) by the use of Non-steroidal anti-inflammatory drugs (NSAIDs) that suppress prostaglandin synthesis~\cite{11}.

The Tumor necrosis factor (TNF$\alpha$) is a cytokine that is induced mainly by activated macrophages and other cell types as well~\cite{12}. It's main role lies in the regulation of immune cells and functionally it is able to induce apoptosis~\cite{13}. Furthermore, it is also interesting to note that TNF$\alpha$ has also been found to interact with NF$\kappa$B1 and SOD [Mn] as well and large amounts of TNF are released in response to lipopolysaccharide products upon stimulus in case of inflammation~\cite{14}.

The cellular tumor antigen or phosphoprotein 53 (TP53) protein is a tumor suppressor protein and it negatively regulates cell division either by stimulation of BAX and FAS antigen expression or by repressing Bcl-2 expression. The main function of TP53 in many tumor types is to induce growth arrest depending on the physiological circumstances~\cite{15}. Further, it also acts as a trans-activator~\cite{16}. As such, TP53 has an important role in preserving stability of the genome by inhibiting mutagens in the genome~\cite{17}.

The Plasminogen Activator, Urokinase (PLAU) enzyme encoded by PLAU gene is a serine protease enzyme. It is primarily involved in the cleavage of zymogen plasminogen to form active form of plasmin~\cite{18}. The production of plasmin results in the thrombolysis of the extracellular matrix of the fibrin lattice in blood clots, which in turn facilitates tissue infiltration leading to metastasis~\cite{19}.

The Nitric oxide synthase, inducible is an enzyme encoded by NOS2 gene. Nitric oxide is a free radical reactive species acting as a mediator in anti-microbial activities. It is generally induced by lipopolysacccharides and cytokines in inflammation~\cite{20}. 

Tissue-type plasminogen activator (PLAT) is a protein involved in the fibrinolysis of blood clots and plays a significant role in cell migration and in various pathological conditions~\cite{21}.

Interleukin-6 (IL6) is a pro-inflammatory cytokine encoded by the IL6 gene secreted by macrophages in response to pathogens. Interleukins are present on the cell surface and plays a significant role in B-cell differentiation into various immune cells such as lymphocytes and monocytes~\cite{22}.

Angiotensin-converting enzyme is an enzyme encoded by the ACE gene that converts Angiotensin I to Angiotensin II leading to vasoconstriction of blood vessels. It mainly inactivates bradykinin which functions as an inflammatory mediator~\cite{23} 

Interleukin-1 Beta (IL1B) may facilitate as a pro-inflammatory cytokine. It mainly induces synthesis of prostaglandins and activates cytokine production accompanied by T-cell / B-cell activation and antibody production~\cite{24}.

Interferon gamma (IFNG) belongs to a class of cytokine that is vital for annate and adaptive immunity against pathogens under type-II class of interferons encoded by the IFNG gene~\cite{25}.

Superoxide dismutase [Cu-Zn] also known as Superoxide dismutase 1 is an enzyme encoded by SOD1 gene that binds to [Cu-Zn] ions and eliminates free radicals produced within the cells. During oxidative stress, SOD1 limits the detrimental effects of reactive oxygen species (ROS) thereby regulating programmed cell death during apoptosis~\cite{26}.

Signal Transducer and activator of transcription 3 (STAT3) is a transcription factor encoded by STAT3 gene. It plays a major role in regulating cell-growth and apoptosis after getting phosphorylated by Mitogen-activated protein kinases (MAPK) / receptor associated Janus kinases (JAK) and ligands such as interferons and interleukins (IL5 and IL6)~\cite{27}.

Methylene tetrahydrofolate reductase (MTHFR) is an enzyme critical in the rate-limiting step in the methyl cycle that is encoded by the MTHFR gene. Methylenetetrahydrofolate reductase deficiency (MTHFRD) is found to be associated with a wide range of autosomal recessive disorders that includes homocysteinuria, homocysteinemia, psychiatric disorders and other neurodegenerative diseases~\cite{28}.

C-C motif chemokine ligand 2 (CCL2) activates the C-C chemokine receptor (CCR2) by acting as a ligand and signals monocytes, dendritic cells and memory T-cells at the inflammation site during infection~\cite{29}. It is present in the plasma membrane of endothelial cells and is mainly induced by macrophages and platelet derived growth factor~\cite{30}.

%\paragraph{}
%
%\begin{acknowledgements}
%The authors would like to thank the Indian Institute of Information Technology, Allahabad (IIIT-A) for providing the infrastructural support and computational facilities. S.C. acknowledges the fellowship received during the course of this study.
%\end{acknowledgements}

% BibTeX users please use one of
%\bibliographystyle{spbasic}      % basic style, author-year citations
%\bibliographystyle{spmpsci}      % mathematics and physical sciences
%\bibliographystyle{spphys}       % APS-like style for physics
%\bibliography{}   % name your BibTeX data base

% Non-BibTeX users please use

\end{document}